# STUDY OF SOME STELLER IRON GROUP FUSION MATERIALS FOR (N,P) REACTIONS

Tarik SIDDIK
College of Basic Education, Department of General Science, Salahaddin University, Erbil, Iraq
e-mail: (taqa@hotmail.com)

Abstract

The excitation functions for (n,p) reactions from reaction threshold to 24 MeV on some important iron (Fe) group target elements ($20 \leq Z \leq 28$) for astrophysical (n, p) reactions such as Si, Ca, Sc, Ti, Cr, Fe, Co and Ni were calculated using TALYS-1.0 nuclear model code. The new calculations on the excitation functions Of $^{28}$Si(n,p)$^{28}$Al, $^{29}$Si(n, p)$^{29}$Al, $^{42}$Ca(n, p)$^{42}$K, $^{45}$Sc(n, p)$^{45}$Ca, $^{46}$Ti(n,p)$^{46}$Sc, $^{52}$Cr(n, p)$^{52}$V, $^{53}$Cr(n, p)$^{53}$V, $^{54}$Fe(n,p)$^{54}$Mn, $^{57}$Fe(n,p)$^{57}$Mn, $^{59}$Co(n, p)$^{59}$Fe, $^{58}$Ni(n, p)$^{58}$Co and $^{60}$Ni(n, p)$^{60}$Co reactions have been carried out up to 24 MeV incident neutron energy.

In these calculations, the pre-equilibrium and equilibrium effects have been investigated. Statistical model calculations, based on the Hauser-Feshbach formalism, have been carried out using the TALYS-1.0 and were compared with existing experimental data as well as with evaluated data files (Experimental Nuclear Reaction Data (EXFOR).According to these calculations, we assume that these model calculations can be applied to some heavy elements, ejected into interstellar medium by dramatic supernova events.

Keywords : Interstellar medium . (n, p) cross–section. Pre-equilibrium reactions.

**1.Introduction**

The stars, are giant nuclear reactors, are born from interstellar matter. Interstellar medium plays a fundamental role in the process of galactic evolution. They deposit energy into the interstellar medium in the form of electromagnetic radiation and stellar winds during their lives. Nuclear fusion is an atomic reaction that fuels stars. When their hydrogen becomes depleted, high mass stars convert He atoms into C and O, followed by the fusion of C and O into Ne, Na, Mg, S and Si. Later reactions transform these elements into Ca, Fe, Ni, Cr, Cu and others. When old stars die, they return some of their matter and energy back into the interstellar medium. More massive stars do this in dramatic supernova events. The matter returned to the interstellar medium is enriched with these heavy elements (such as Fe, Si, Ca, Sc, Ti, Cr and Ni) produced by the nuclear burning in the stars' interior and in processes occurring during the explosions. This enriched ejecta becomes the material for future generations of stars which behave rather differently due to the presence of the metals. The process is not 100% efficient; some of the matter remains locked up forever in compact objects such as white dwarfs, neutron stars and black holes. Therefore, the composition of a galaxy slowly proceeds from all interstellar medium to no Interstellar medium. The chemical evolution of the galaxy results from this cycle of stellar birth, death, and enriched rebirth. The growth in the quantity of heavy elements allows, along the way, the possible formation of planets, rocks, and in at least one case living organisms. Reactions of interest often concern unstable or even exotic (neutron-rich, neutron-deficient, superheavy) species for which no experimental data exist. Given applications (in particular, nuclear astrophysics and accelerator-driven systems) involve a large number (thousands) of unstable nuclei for which many different properties have to be determined [1-4].

Finally Most of the (n, p) reaction cross-section data are available at about 14–15 MeV. Significant amount of new nuclear data is required new nuclear energy system in extended energy region. Limitations of the activation technique to the stable products, unsuitable half lives and uncertain decay schemes account for the lack of the data. The lack of experimental data has to be compensated by the development of reliable calculation methods [5].

Reaction rates for medium or heavy nuclei, applied in astrophysical applications, have been calculated until now on the basis of the Hauser-Feshbach (HF) statistical model, adopting simplified schemes to estimate the capture reaction cross section of a given target nucleus, not only in its ground state but also in the different thermally populated states of the stellar plasma at a given temperature [6-8].

In the present work, the new calculations on the excitation functions of $^{28}$Si(n,p)$^{28}$Al, $^{29}$Si(n, p)$^{29}$Al, $^{42}$Ca(n, p)$^{42}$K, $^{45}$Sc(n, p)$^{45}$Ca, $^{46}$Ti(n,p)$^{46}$Sc, $^{52}$Cr(n, p)$^{52}$V, $^{53}$Cr(n, p)$^{53}$V, $^{54}$Fe(n,p)$^{54}$Mn, $^{57}$Fe(n, p)$^{57}$Mn, $^{59}$Co(n, p)$^{59}$Fe, $^{58}$Ni(n, p)$^{58}$Co and $^{60}$Ni(n, p)$^{60}$Co reactions have been carried out up to 24 MeV incident neutron energy.

## 2. Models

An outline of the general theory and modeling of nuclear reactions can be given in many ways. A common classification is in terms of time scales: short reaction times are associated with direct reactions and long reaction times with compound nucleus processes. At intermediate time scales, pre-equilibrium processes occur. An alternative, more or less equivalent, classification can be given with the number of intranuclear collisions, which is one or two for direct reactions, a few for pre-equilibrium reactions and many for compound reactions, respectively. As a consequence, the coupling between the incident and outgoing channels decreases with the number of collisions and the statistical nature of the nuclear reaction theories increases with the number of collisions. This distinction between nuclear reaction mechanisms can be obtained in a more formal way by means of a proper division of the nuclear wave functions into open and closed configurations, as detailed for example by Feshbach's many contributions to the field. When discussing nuclear reactions in the context of a computer code, a different starting point is more appropriate. A particle incident on a target nucleus will induce several binary reactions which are described by the various competing reaction mechanisms that were mentioned above. The end products of the binary reaction are the emitted particle and the corresponding recoiling residual nucleus. In general this is, however, not the end of the process. A total nuclear reaction may involve a whole sequence of residual nuclei, especially at higher energies, resulting from multiple particle emission. All these residual nuclides have their own separation energies, optical model parameters, level densities, fission barriers, gamma strength functions, etc., that must properly be taken into account along the reaction chain. The implementation of this entire reaction chain forms the backbone of TALYS. The program has been written in a way that enables a clear and easy inclusion of all possible nuclear model ingredients for any number of nuclides in the reaction chain. Of course, in this whole chain the target and primary compound nucleus have a special status, since they are subject to all reaction mechanisms, i.e. direct, pre-equilibrium, compound and fission and, at low incident energies, width fluctuation corrections in compound nucleus decay. Also, at incident energies below a few MeV, only binary reactions take place and the target and compound nucleus are often the only two nuclei involved in the whole reaction. Historically, it is for the binary reactions that most of the theoretical methods have been developed and refined, mainly because their validity, and their relation with nuclear structure, could best be tested with exclusive measurements.
The nuclear models that are included in TALYS, generally can be categorized into optical, direct, pre-equilibrium, compound and fission models, all driven by a comprehensive database of nuclear structure and model parameters [9,10].

## 2.1. Optical model

The central assumption underlying the optical model is that the complicated interaction between an incident particle and a nucleus can be represented by a complex mean-field potential, which divides the reaction flux into a part covering shape elastic scattering and a part describing all competing non-elastic channels. Solving the Schrodinger equation numerically with this complex potential yields a wealth of valuable information. First, it returns a prediction for the basic observables, namely the elastic angular distribution and polarization, the reaction and total cross section and, for low energies, the *s*,*p*-wave strength functions and the potential scattering radius R'. The essential value of a good optical model is that it can reliably predict these quantities for energies and nuclides for which no measurements exist.
Also, the quality of the not directly observable quantities that are provided by the optical model has an equally important impact on the evaluation of the various reaction channels. Well-known examples are transmission coefficients, for compound nucleus and multi-step compound decay, and the distorted wave functions that are used for direct inelastic reactions and for transitions to the continuum that describe statistical multi-step direct reactions. Also, the reaction cross sections that are calculated with the optical model are crucial for the semi-classical pre-equilibrium models. All optical model calculations are performed by ECIS-06, which is used as a subroutine in TALYS. The default optical-model potentials (OMP) used in TALYS are the local and global parameterisations for neutrons and protons of [11-14].

## 2.3. Direct reactions

Various models for direct reactions are included in the program: DWBA for (near-)spherical nuclides, coupled-channels for deformed nuclides, the weak-coupling model for odd nuclei, and also a giant resonance contribution in the continuum. In all cases, TALYS drives the ECIS-06 code to perform the calculations. The results are presented as discrete state cross sections and angular distributions, or as contributions to the continuum.

## 2.4. Pre-equilibrium reactions

It is now well-known that the separation of nuclear reaction mechanisms into direct and compound is too simplistic. Furthermore, the measured angular distributions in the region between direct and compound are anisotropic, indicating the existence of a memory-preserving, direct-like reaction process. Apparently, as an intermediate between the two extremes, there exists a reaction type that embodies both direct- and compound-like features. These reactions are referred to as pre-equilibrium, precompound or, when discussed in a quantum-mechanical context, multi-step processes. Pre-equilibrium emission takes place after the first stage of the reaction but long before statistical equilibrium of the compound nucleus is attained. It is imagined that the incident particle step-by-step creates more complex states in the compound system and gradually loses its memory of the initial energy and direction. Pre-equilibrium processes cover a sizable part of the reaction cross section for incident energies between 10 and (at least) 200 MeV. Pre-equilibrium reactions have been modeled both classically and quantum-mechanically and both are included in TALYS.

## 2.5. Compound Nucleus reactions

The term compound nucleus reaction is commonly used for two different processes: (i) capture of the projectile in the target nucleus to form a compound nucleus,which subsequently emits a particle or gamma; (ii) multiple emission from the chain of excited residual nuclides following the binary reaction. Both are included in TALYS. At low incident energy (i) plays an important role. It differs from (ii) at two important points: (a) the presence of width-fluctuation corrections and (b) nonisotropic,though still symmetric, angular distributions. It is calculated with the Hauser-Feshbach formalism including width fluctuation corrections (WFC). The WFC factor accounts for the correlations that exist between the incident and outgoing waves. From a qualitative point of view, these correlations enhance the elastic channel and accordingly decrease the other open channels. In general, the WFC factor may be calculated using three different expressions, which have all been implemented in TALYS: The HRTW model, the Moldauer model, and the model using the Gaussian Orthogonal Ensemble (GOE).
The Moldauer model is confirmed as the best default choice. All WFC models are generalized to include continuum particle emission, gamma-ray competition, and fission. Gamma-ray coefficients are modeled with Kopecky-Uhl's generalized Lorentzian and the appropriate giant-dipole resonance parameters. Besides cross sections, compound angular distributions are calculated using Blatt-Biedenharn coupling factors, again within a full Hauser-Feshbach expression with WFC .For multiple emission, the whole reaction chain is followed by depleting each [nucleus-excitation energy spin-parity] bin with particle, gamma, or fission decay until all channels are closed. In the process, all particle and residual production cross sections are accumulated to their final values. Non-equidistant energy grids in this decay scheme ensure enough precision in the compound evaporation peaks [15-20].

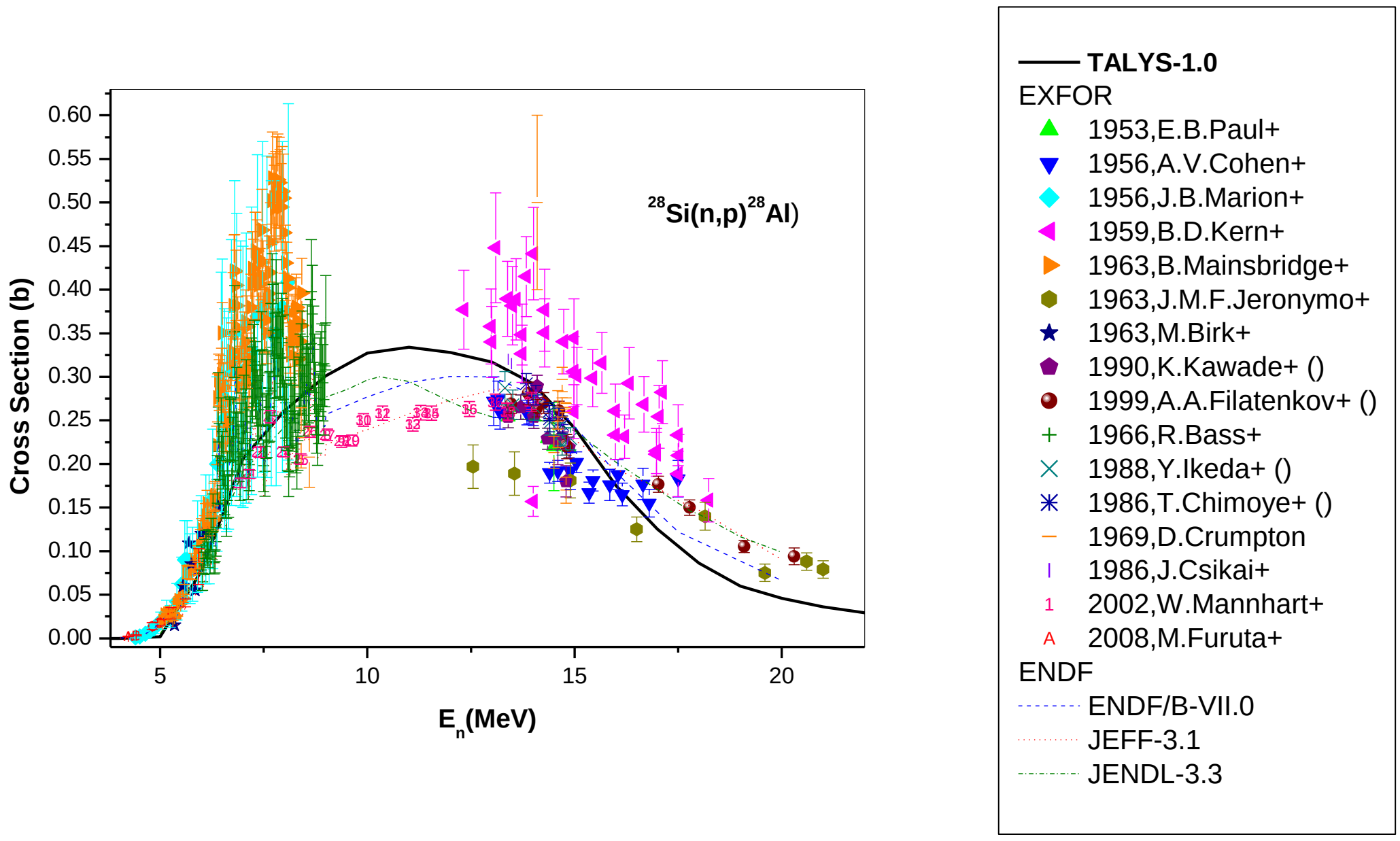

Fig. 1 The comparison of calculated excitation function using TALS-1.2 of $^{28}$Si(n,p)$^{28}$Al reaction with available experimental values and evaluated nuclear data files ENDF/B-VII.0, JENDL-3.3 and JEFF3.1.The values reported in Ref. (21)

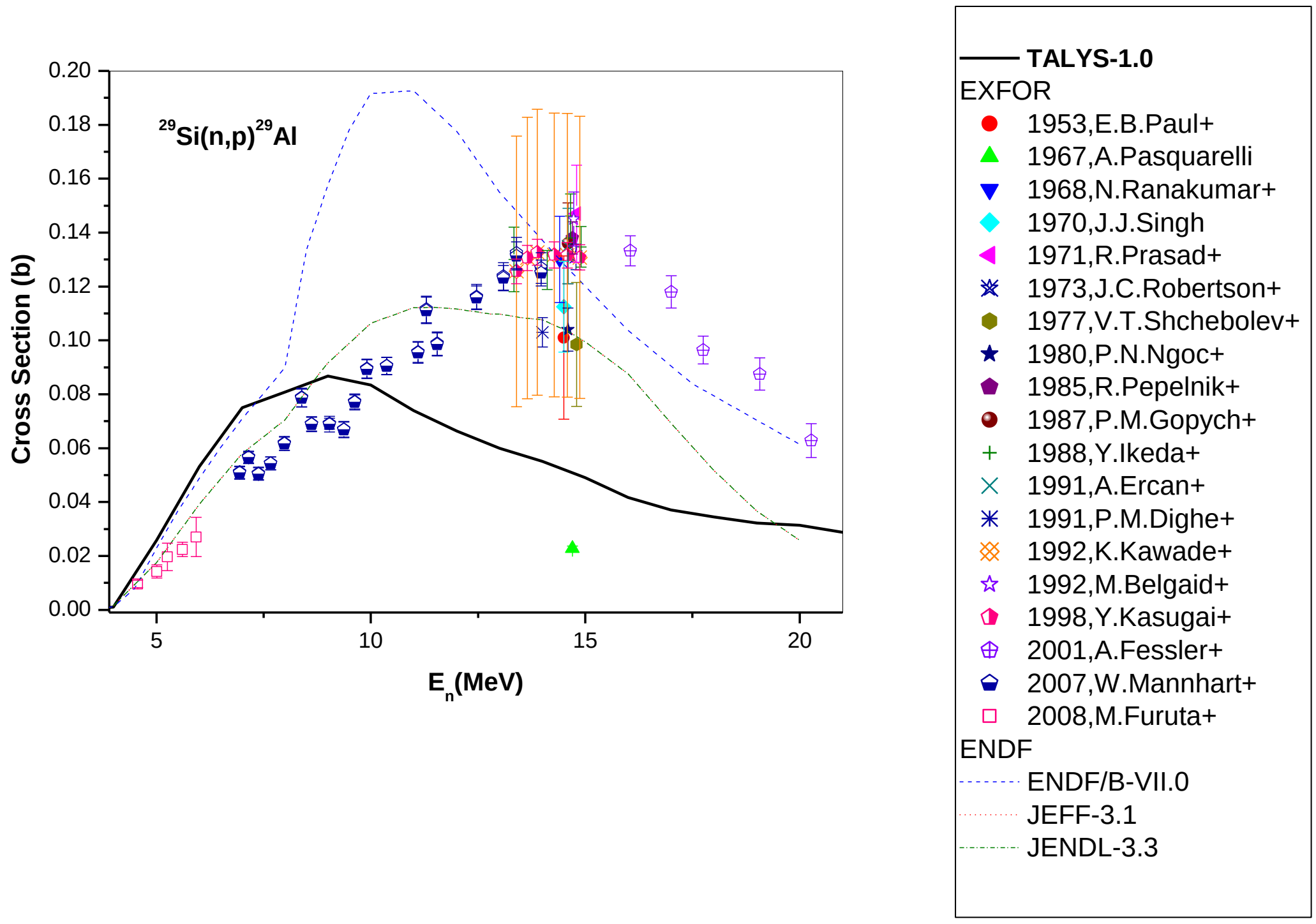

Fig. 2 The comparison of calculated excitation function using TALS-1.2 of $^{29}Si(n,p)^{29}Al$ reaction with available experimental values and evaluated nuclear data files ENDF/B-VII.0, JENDL-3.3 and JEFF3.1The values reported in Ref. (21)

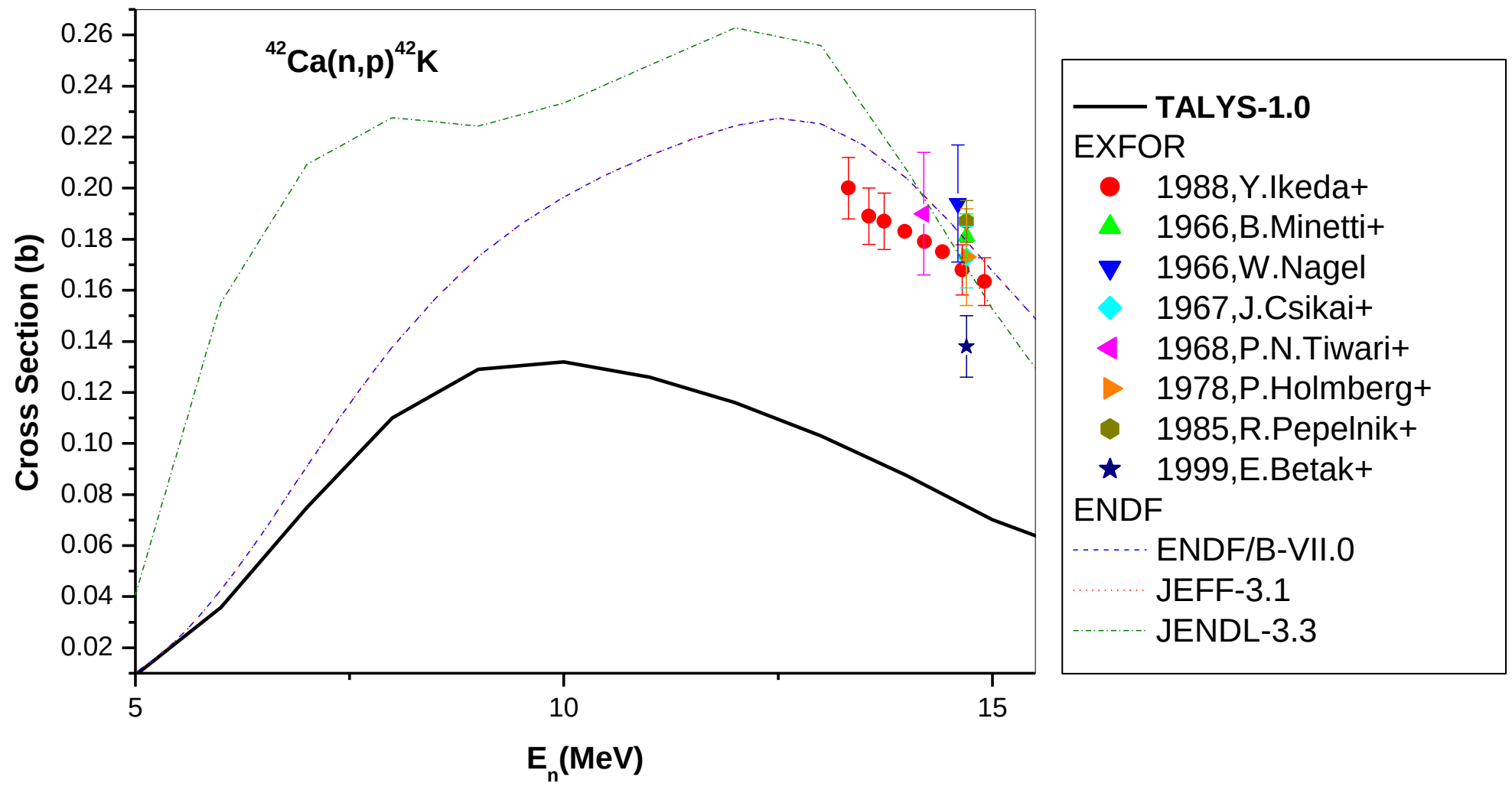

Fig. 3 The comparison of calculated excitation function using TALS-1.2 of $^{42}Ca(n,p)^{42}K$ reaction with available experimental values and evaluated nuclear data files ENDF/B-VII.0, JENDL-3.3 and JEFF3.1The values reported in Ref. (21)

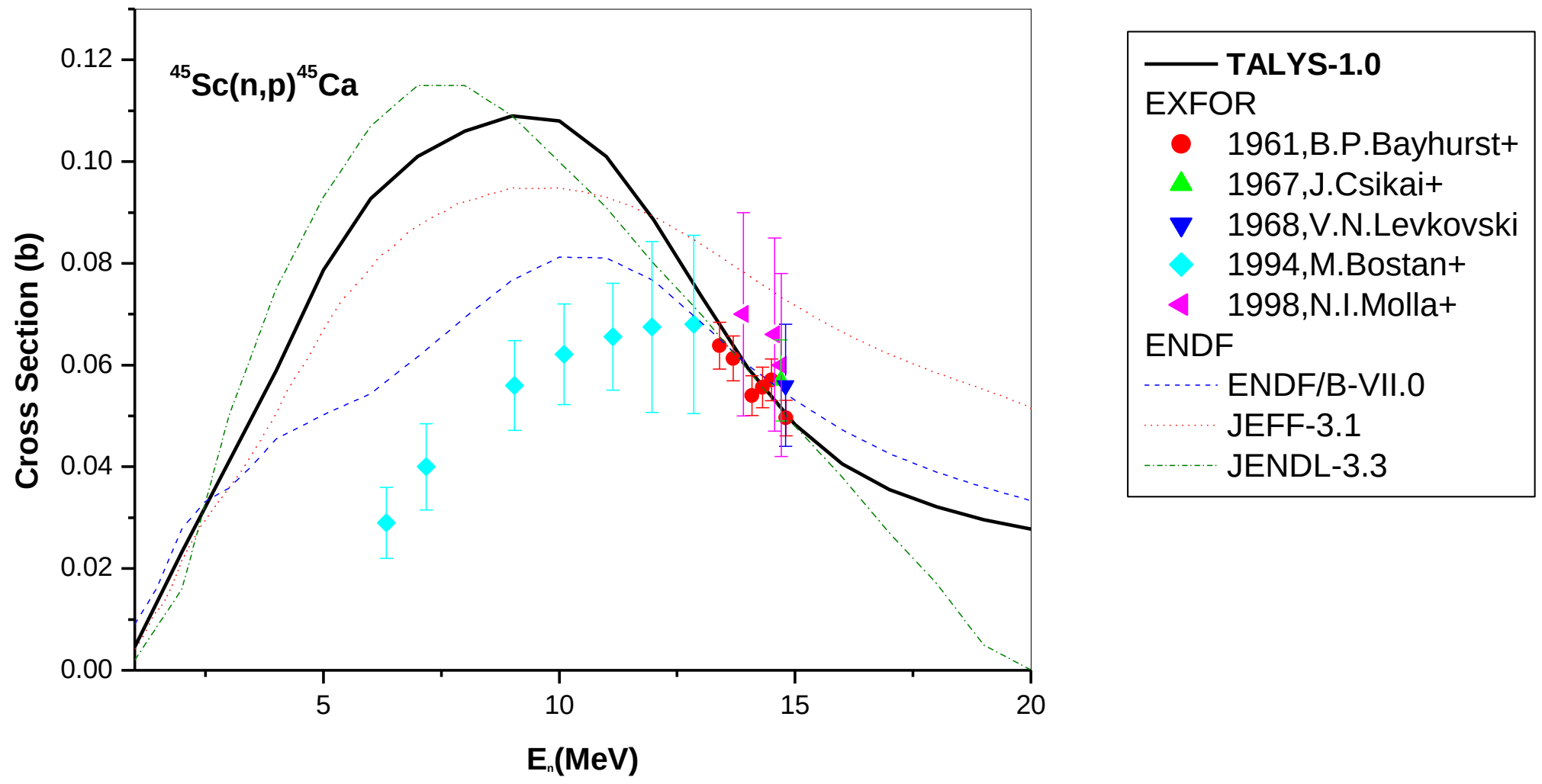

Fig. 4 The comparison of calculated excitation function using TALS-1.2 of $^{45}Sc(n,p)\ ^{45}Ca$ reaction with available experimental values and evaluated nuclear data files ENDF/B-VII.0, JENDL-3.3 and JEFF3.1The values reported in Ref. (21)

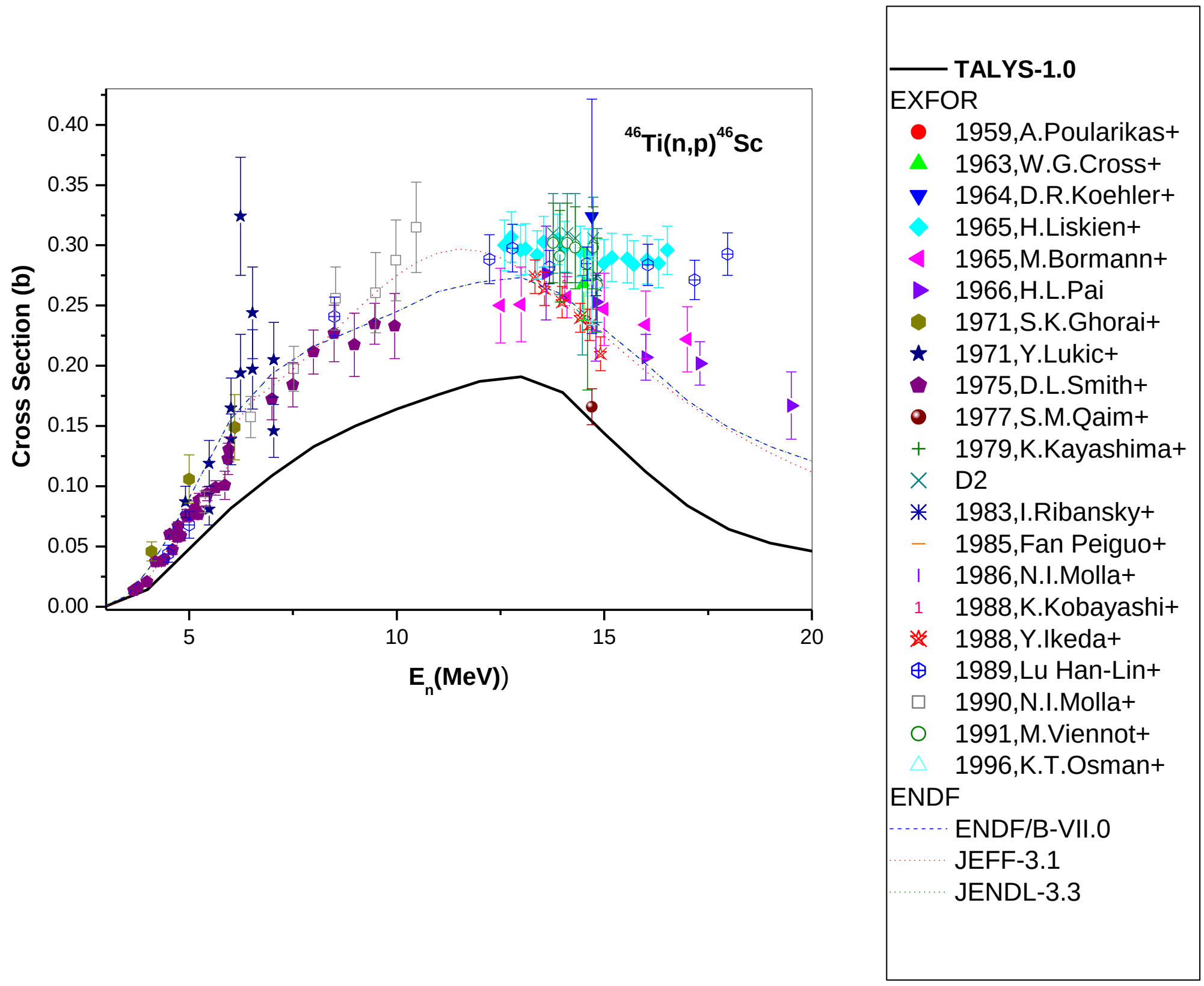

Fig. 5 The comparison of calculated excitation function using TALS-1.2 of $^{48}$Ti(n,p) $^{48}$Sc reaction with available experimental values and evaluated nuclear data files ENDF/B-VII.0, JENDL-3.3 and JEFF3.1.The values reported in Ref. (21)

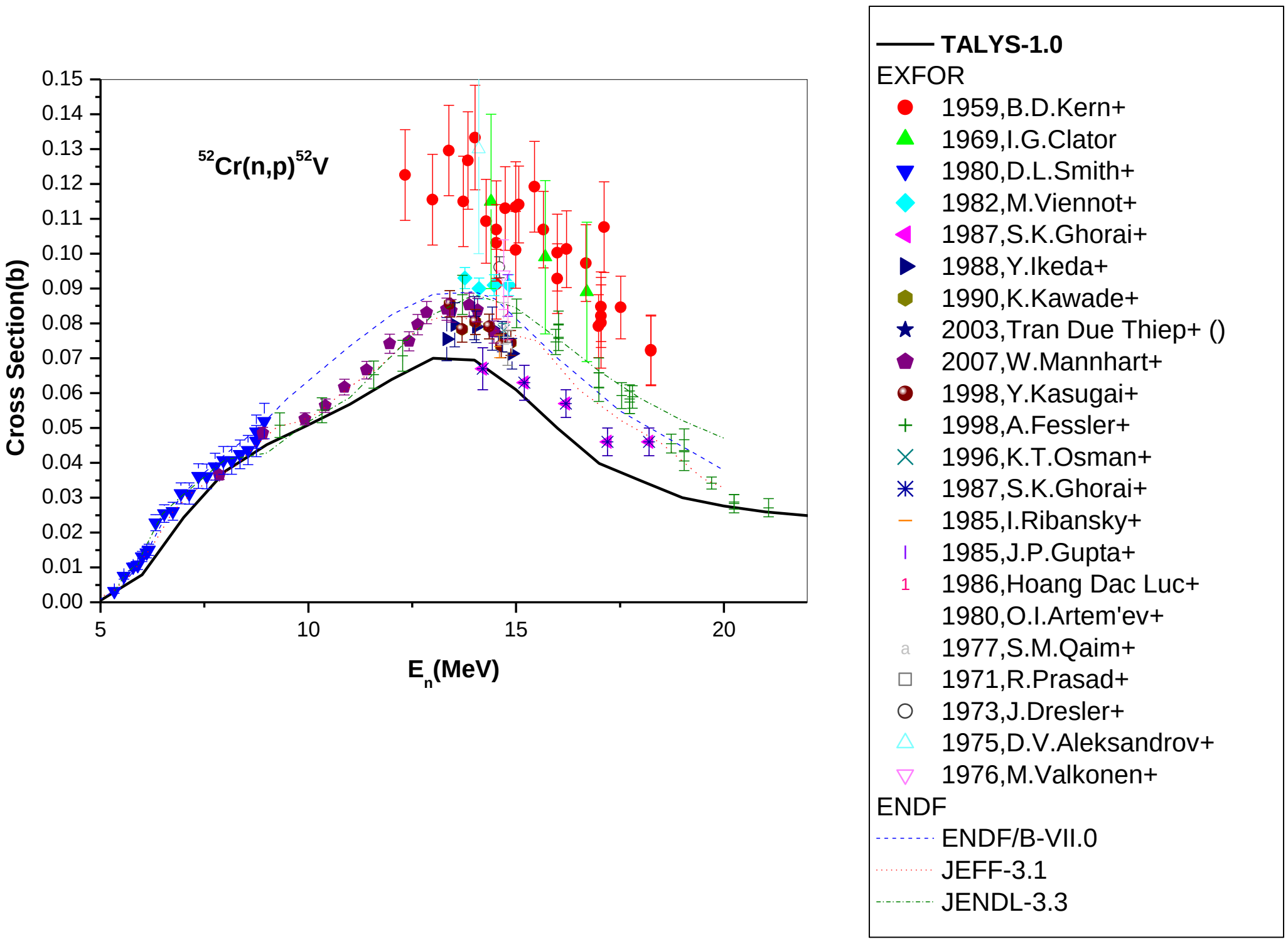

Fig.6 The comparison of calculated excitation function using TALS-1.2 of $^{52}$Cr(n,p) $^{52}$V reaction with available experimental values and evaluated nuclear data files ENDF/B-VII.0, JENDL-3.3 and JEFF3.1.The values reported in Ref. (21)

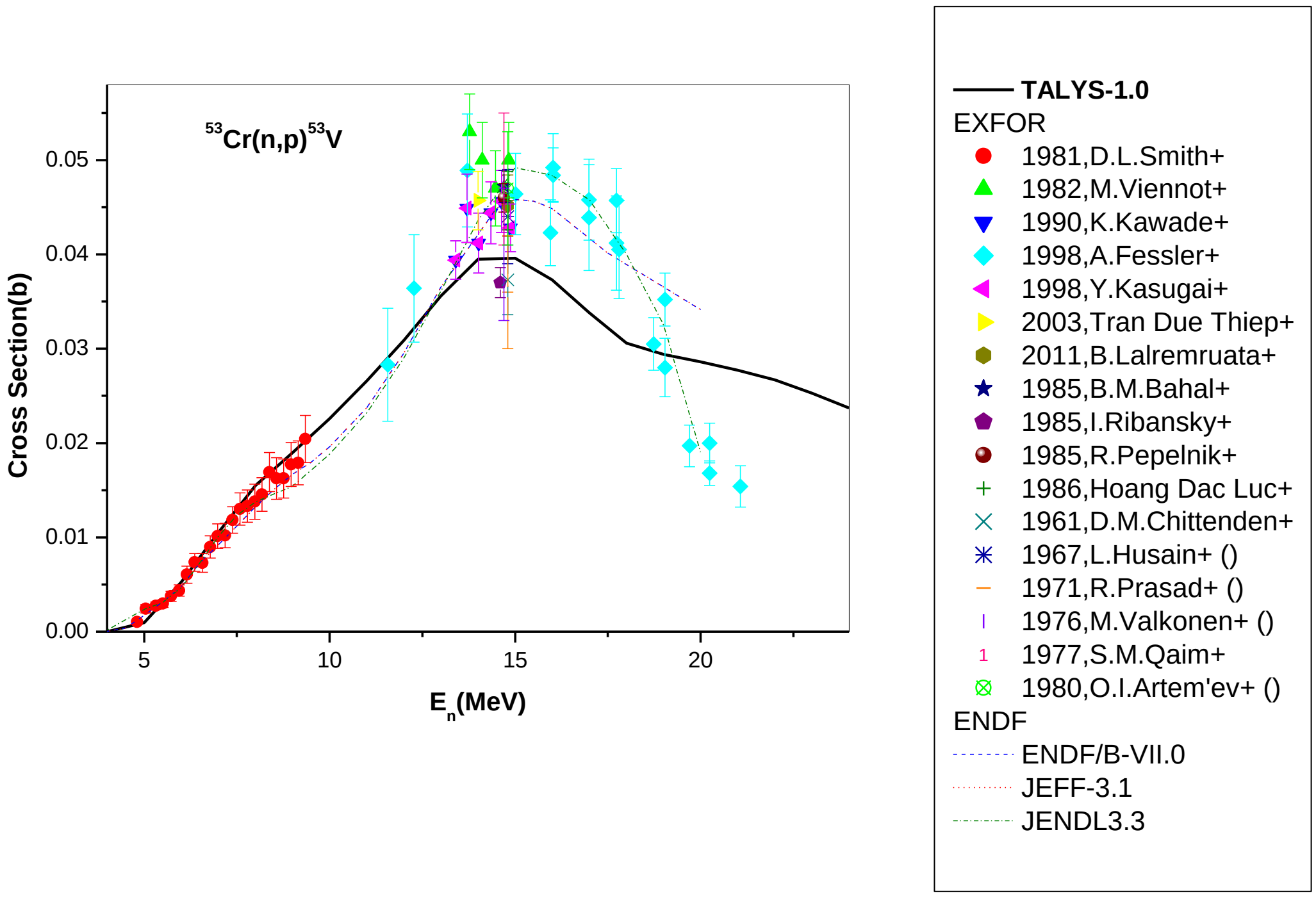

Fig. 7 The comparison of calculated excitation function using TALS-1.2 of $^{53}$Cr(n,p) $^{53}$V reaction with available experimental values and evaluated nuclear data files ENDF/B-VII.0, JENDL-3.3 and JEFF3.1.The values reported in Ref. (21)

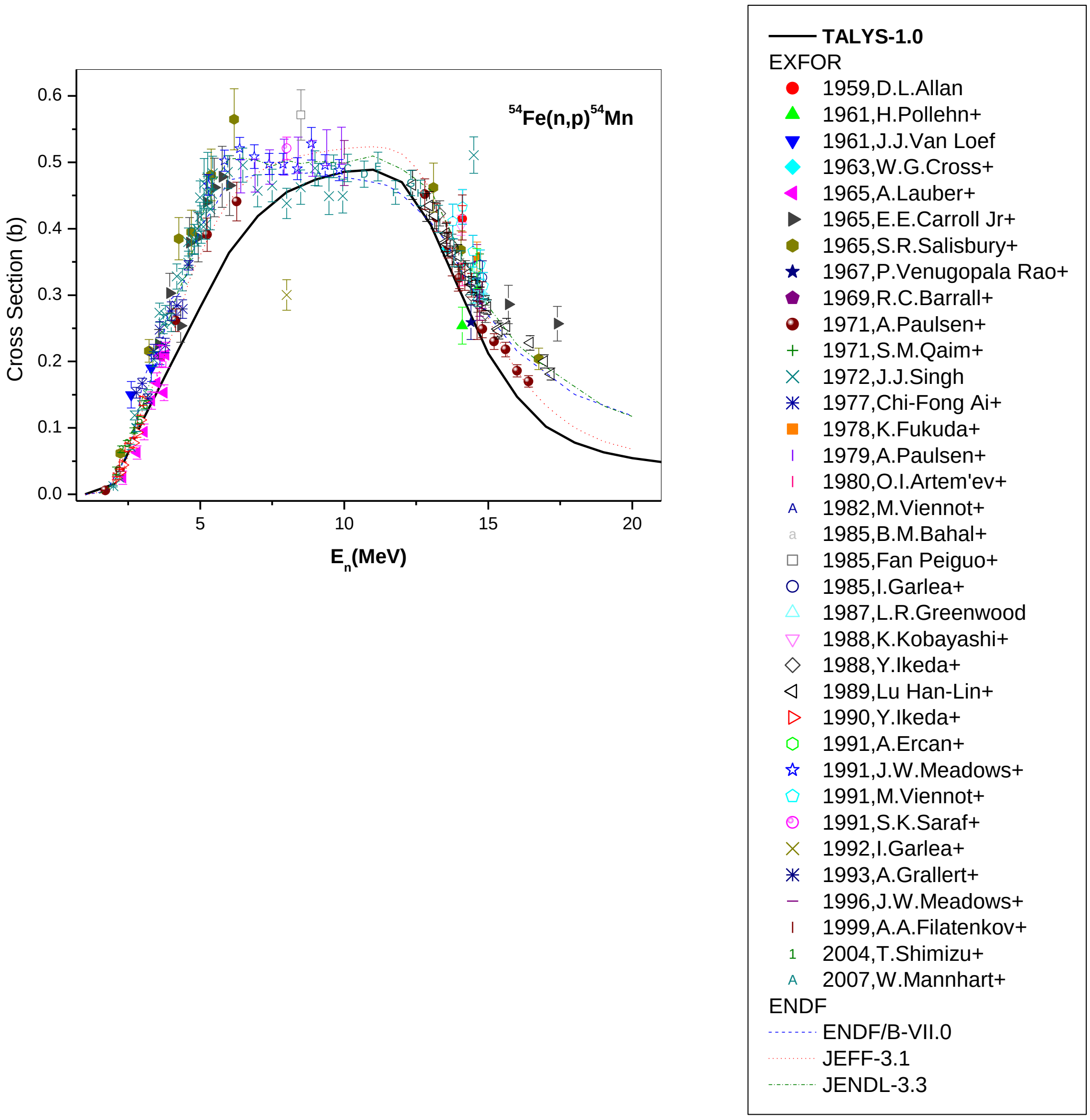

Fig. 8 The comparison of calculated excitation function using TALS-1.2 of $^{54}$Fe(n,p) $^{54}$Mn reaction with available experimental values and evaluated nuclear data files ENDF/B-VII.0, JENDL-3.3 and JEFF3.1.The values reported in Ref. (21)

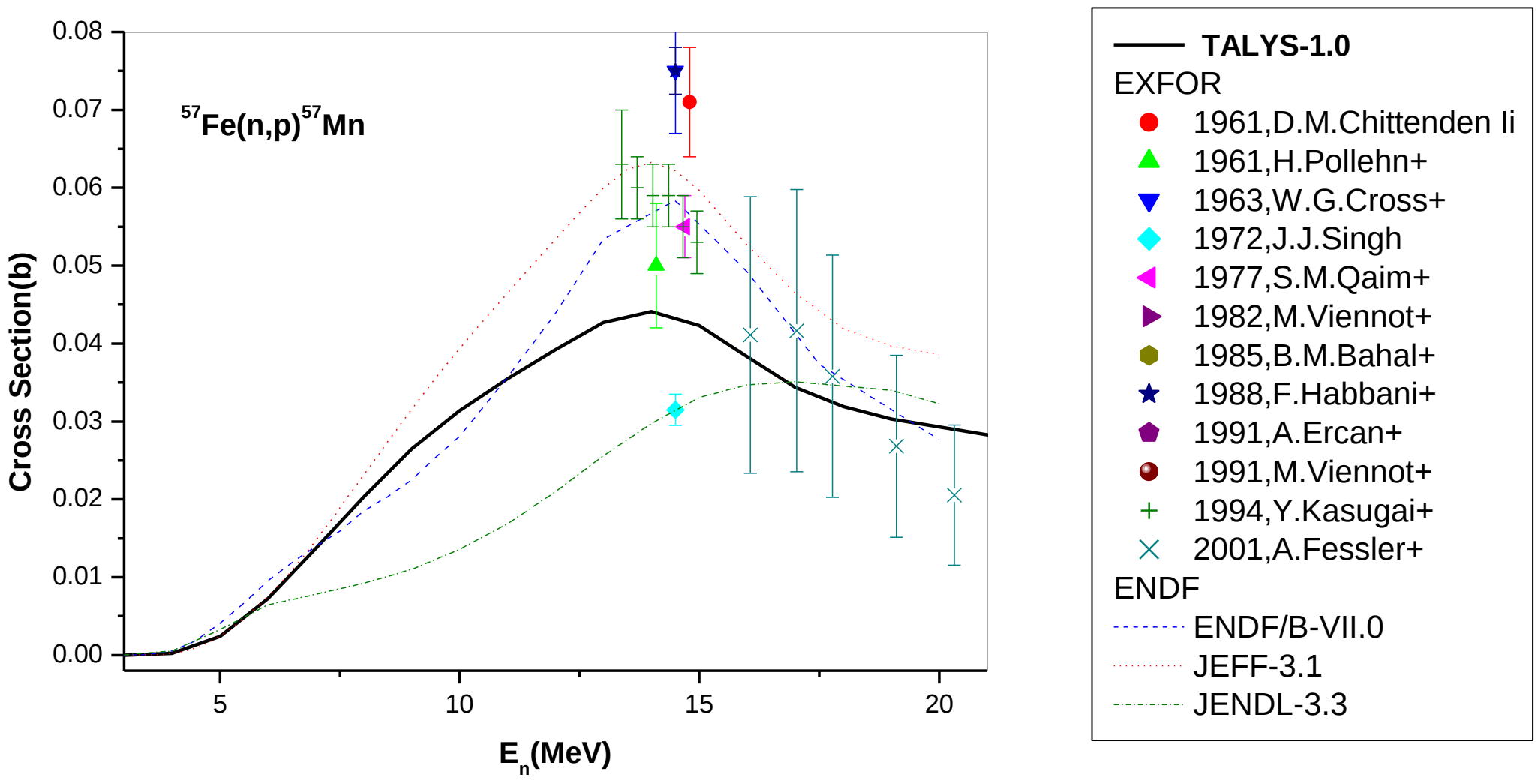

Fig.9 The comparison of calculated excitation function using TALS-1.2 of $^{57}Fe(n,p)$ $^{57}Mn$ reaction with available experimental values and evaluated nuclear data files ENDF/B-VII.0, JENDL-3.3 and JEFF3.1.The values reported in Ref. (21)

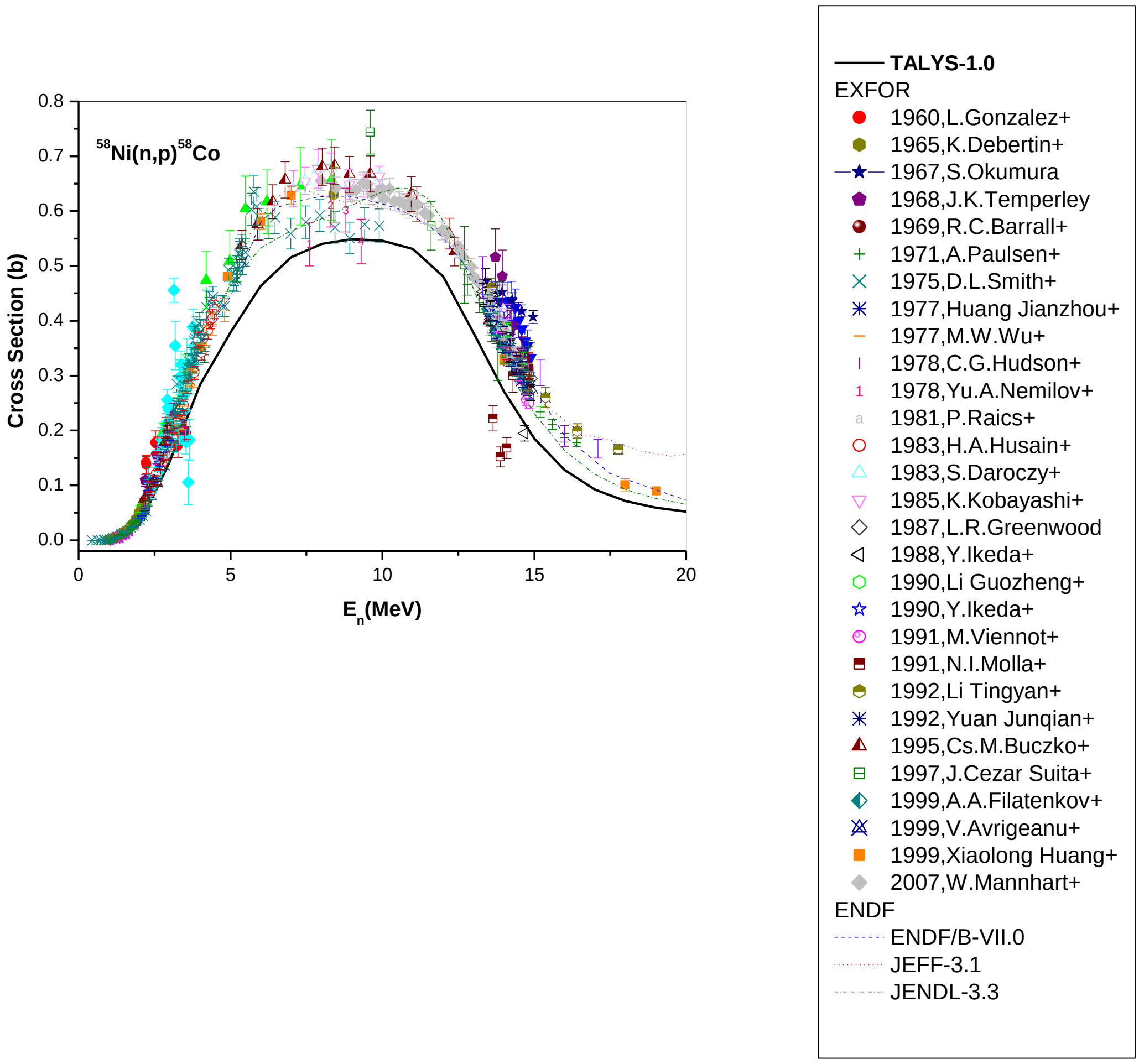

Fig. 10 The comparison of calculated excitation function using TALS-1.2 of $^{58}Ni(n,p)$ $^{58}Co$ reaction with available experimental values and evaluated nuclear data files ENDF/B-VII.0, JENDL-3.3 and JEFF3.1.The values reported in Ref. (21)

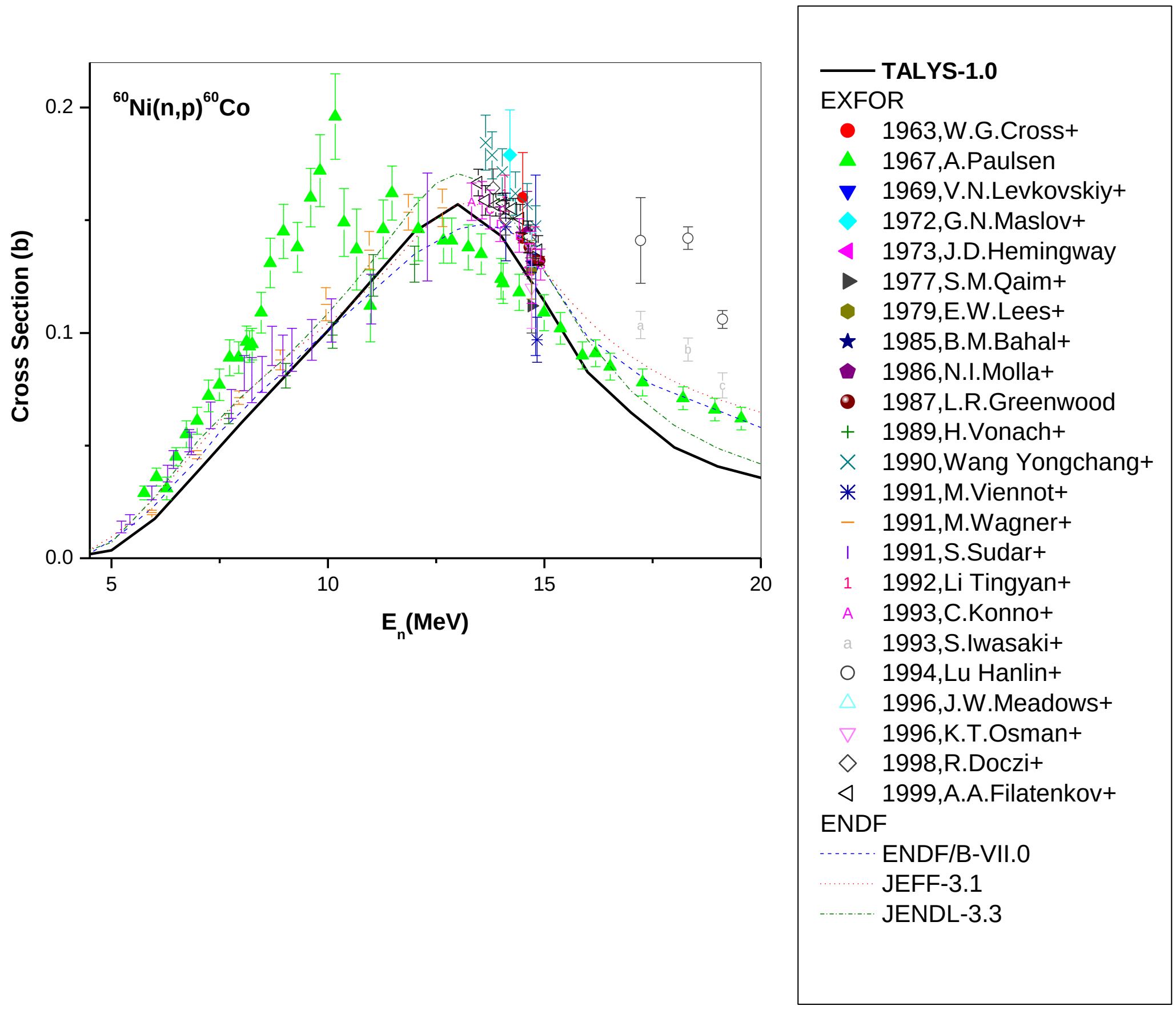

Fig. 11 The comparison of calculated excitation function using TALS-1.2 of $^{60}$Ni(n,p) $^{60}$Co reaction with available experimental values and evaluated nuclear data files ENDF/B-VII.0, JENDL-3.3 and JEFF3.1.The values reported in Ref. (21)

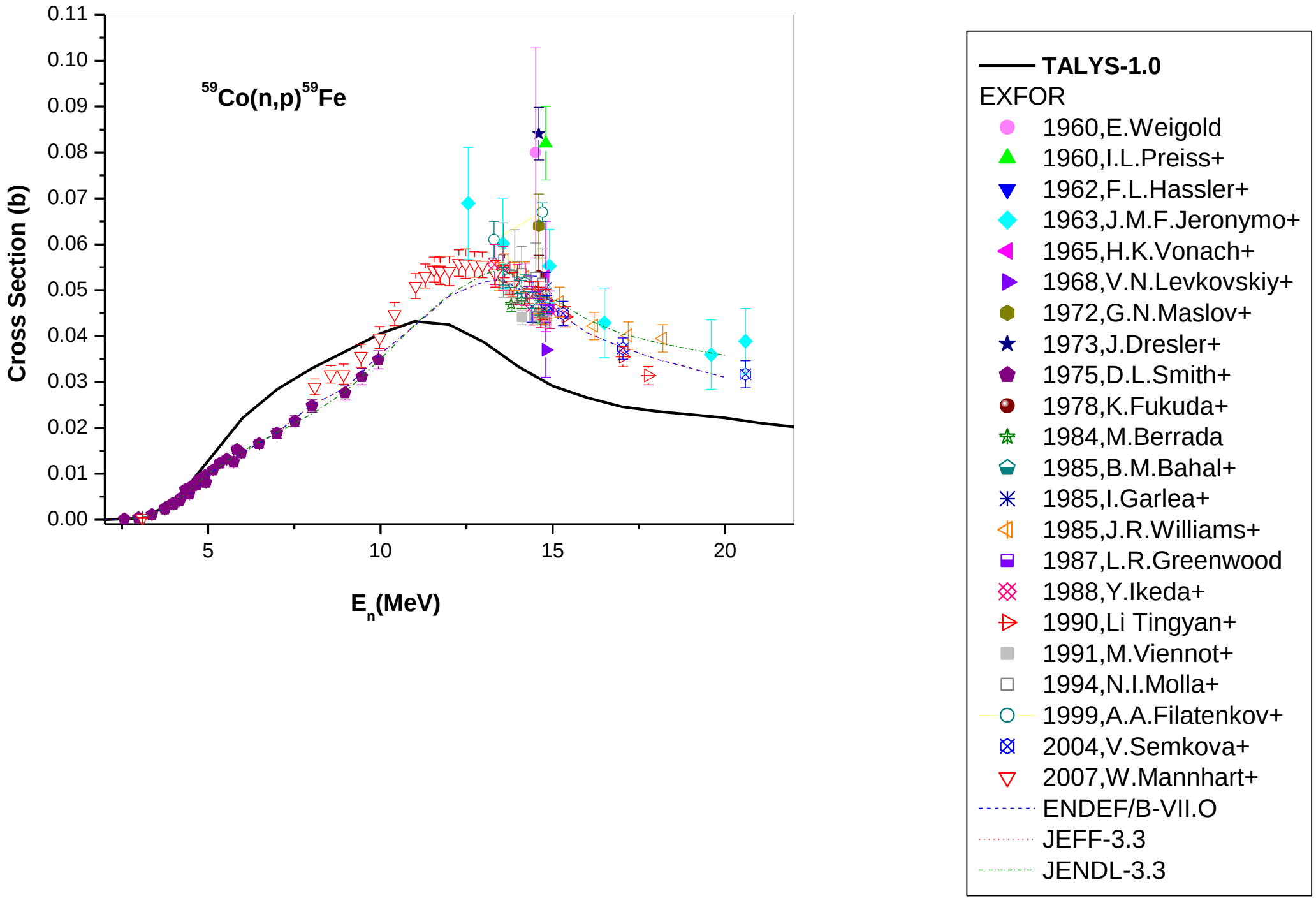

Fig. 12 The comparison of calculated excitation function using TALS-1.2 of $^{59}Co(n,p)^{59}Fe$ reaction with available experimental values and evaluated nuclear data files ENDF/B-VII.0, JENDL-3.3 and JEFF3.1.The values reported in Ref. (21)

## 3. Calculations and analysis :

TALYS is a nuclear reaction program created at NRG Petten, the Netherlands, and CEA Bruyères-le-Châtel,France. The idea to make TALYS was born in 1998, it also provides a complete and accurate simulation of nuclear reactions in the 1 keV - 200 MeV energy range. There are two main purposes of TALYS, which are strongly connected. First, it is a nuclear physics tool that can be used for the analysis of nuclear-reaction experiments.The interplay between experiment and theory gives us insight in the fundamental interaction between particles and nuclei, and precise measurements enable us to constrain our models and their parameters. In return,when the resulting nuclear models are believed to have sufficient predictive power, the prediction can even give an indication of the reliability of measurements. After the nuclear physics stage comes the second function of TALYS, namely as a nuclear data tool: After fine-tuning the adjustable parameters of the various reaction models using available experimental data, TALYS can generate nuclear data for all open reaction channels,interpolating between and extrapolating beyond experimental data, on a user-defined energy and angle grid beyond the resonance region. The associated nucleardata libraries that can be constructed provide essential information for existing and new nuclear technologies.Important applications that rely directly on the output of nuclear-reaction simulation software like TALYS are conventional and innovative power reactors (GENIV),accelerator-driven systems, and transmutation of radioactive waste, fusion reactors, homeland security, medical-isotope production, and radiotherapy, oil-well logging, geophysics, and astrophysics.

In present paper,the excitation functions for (n,p) reactions from reaction threshold to 24 MeV on some important iron (Fe) group target elements (20 ≤ Z ≤28) for astrophysical (n, p) reactions such as Si, Ca, Sc, Ti, Cr, Fe, Co and Ni were calculated using TALYS-1.0 nuclear model code. The new calculations on the excitation functions Of $^{28}Si(n,p)^{28}Al$, $^{29}Si(n, p)^{29}Al$, $^{42}Ca(n, p)^{42}K$, $^{45}Sc(n, p)^{45}Ca$, $^{46}Ti(n,p)^{46}Sc$, $^{52}Cr(n, p)^{52}V$, $^{53}Cr(n, p)^{53}V$, $^{54}Fe(n,p)^{54}Mn$, $^{57}Fe(n,p)^{57}Mn$, $^{59}Co(n, p)^{59}Fe$, $^{58}Ni(n, p)^{58}Co$ and $^{60}Ni(n, p)^{60}Co$ reactions have been carried out up to 24 MeV incident neutron energy. The calculated results have been also compared with available experimental data in the literature and with ENDF/B-VII, T=300k; JENDL-3.3, T=300k and JEFF3.1, T=300k evaluated libraries . A reasonable agreement with experimental and theoretical excitation functions was obtained. The results can be summarized and conclude as follows:

3.1. The $^{28}Si(n,p)^{28}Al$ reaction cross section

The experimental data points are between 1-24 MeV. ENDF/B-VI, JENDL-3.3 and JEFF3.1 files are in agreement with each other. There is good agreement between the cross-section calculated with TALYS-1.0 and the experimental data.

3.2. The $^{29}Si(n,p)^{29}Al$ reaction cross section

The experimental data points are between 1-24 MeV. ENDF/B-VI and JENDL-3.3 files are not in agreement with each other. There is good agreement between the cross-section calculated with TALYS-1.0 and the experimental data in low energy. The calculations between 11-24MeV are not in good agreement with experimental data for $^{29}Si(n,p)^{29}Al$ reaction .

3.3. The $^{42}Ca(n,p)^{42}K$ reaction cross section

The experimental points are between 13 - 15 MeV, ENDF/B-VI and JENDL-3.3 files are in agreement with each other. It could not be said that the calculation are in good agreement with experimental data for $^{42}Ca(n,p)^{42}K$ reaction.

3.4. The $^{46}Sc(n,p)^{46}Ca$ reaction cross section

The experimental points are between 6 - 15 MeV, ENDF/B-VI and JENDL-3.3 files are in agreement with each other. The calculations between 12-15 MeV are in good agreement with experimental data for $^{46}Sc(n,p)^{46}Ca$ reaction .

3.5. The $^{52}Cr(n,p)^{52}V$ reaction cross section

The experimental data points are between 5-21 MeV. ENDF/B-VI, JENDL-3.3 and JEFF3.1 files are in agreement with each other. There is good agreement between the cross-section calculated with TALYS-1.0 and the experimental data.

3.6. The $^{53}Cr(n,p)^{53}V$ reaction cross section

The experimental data points are between 4-21 MeV. ENDF/B-VI, JENDL-3.3 and JEFF3.1 files are in agreement with each other. There is good agreement between the cross-section calculated with TALYS-1.0 and the experimental data.

3.7. The $^{54}Fe(n,p)^{54}Mn$ reaction cross section

The experimental data points are between 1-18 MeV. ENDF/B-VI, JENDL-3.3 and JEFF3.1 files are in agreement with each other. There is good agreement between the cross-section calculated with TALYS-1.0 and the experimental data.

3.8. The $^{57}Fe(n,p)^{57}Mn$ reaction cross section

The experimental data points are between 14-21 MeV. ENDF/B-VI and JEFF3.1 files are in agreement with each other. There is good agreement between the cross-section calculated with TALYS-1.0 and the experimental data.

3.9. The $^{58}Ni(n,p)^{58}Co$ reaction cross section

The experimental data points are between 1-18 MeV. ENDF/B-VI, JENDL-3.3 and JEFF3.1 files are in agreement with each other. There is good agreement between the cross-section calculated with TALYS-1.0 and the experimental data.

3.10. The $^{60}Ni(n,p)^{60}Co$ reaction cross section

The experimental data points are between 6-18 MeV. ENDF/B-VI, JENDL-3.3 and JEFF3.1 files are in agreement with each other. There is excellent agreement between the cross-section calculated with TALYS-1.0 and the experimental data.

3.11. The $^{59}Co(n,p)^{59}Fe$ reaction cross section

The experimental data points are between 1-21 MeV. ENDF/B-VI, JENDL-3.3 and JEFF3.1 files are in agreement with each other. There is excellent agreement between the cross-section calculated with TALYS-1.0 and the experimental data in low energy.

References:

[1] Reynolds R.J., Science, 277,1446(1997).
[2] Ferriere K.M.Rev.Mod.phys.,73,1031(2001).
[3] Spitzer L.Jr. Physical Processes in the Interstellar Medium, New York :Willey-Interscience(1978).
[4] Dopita M.A.,Sutherland R.S. Astrophysics of the Diffuse Universe, Heidelberg: Springer, kapitel 11(Bibliothek des Astronomischen Institutus)( 2002).
[5] Tarik S.RESHID , J.Fusion Energy Volume 32, Issue 2(2013),Page 164-170.
[6] Eyyup.TEL, J.Fusion Energy 29:332-336(2010).
[7] Hauser, W., & Feshbach, H.Phys. Rev., 87, 366.( 1952).
[8] Eyyup.TEL et al., J.Fusion Energy 27(3), 188(2008).
[9] A.J.Koning et al. User Manual of the Code TALYS-1.0 (2007)
[10] A.J. Koning, S. Hilaire and M.C. Duijvestijn, .TALYS: Comprehensive nuclear reaction modeling ., *Proceedings of the International Conference on Nuclear Data for Science and Technology -ND2004*, Santa Fe, USA, p. 1154 (2005).
[11] A.J. Koning and M.C. Duijvestijn, .New nuclear data evaluations for Ge isotopes. Nucl. Instr.Meth. B248, 197 (2006).
[12] J. Raynal, *Notes on ECIS94*, CEA Saclay Report No. CEA-N-2772, (1994).
[13] A.J. Koning and J.P. Delaroche, Nucl. Phys. **A713**, 231(2003).
[14] S. Hilaire, Ch. Lagrange and A.J. Koning, Ann. Phys.**306**, 209 (2003).
[15] J.M. Akkermans and H. Gruppelaar, Phys. Lett. **157B**,95 (1985).
[16] E. Sh Soukhovitskii, S. Chiba, J.-Y. Lee, O. Iwamoto and T. Fukahori, J. Phys. G: Nucl. Part. Phys.30, 905 (2004).
[17] Tarik S.RESHID et al. "Calculation of Neutron-Induced Cross Section for the Stable Isotopes of Pb" Proceedings of International Conference on Nuclear Data and Science Application,14-17 October ,Ankara,Turkey (2008).
[18] H. Gruppelaar, P. Nagel, and P.E. Hodgson, Riv. Nuovo Cimento **9**, No. 7, 1 (1986).
[19] E. Gadioli and P.E. Hodgson, *Pre-equilibrium nuclear reactions*, Oxford Univ. Press (1992).
[20] A.J. Koning and M.C. Duijvestijn, Nucl. Phys. **A744** (2004) 15.
[21] Brookhaven National Laboratory, National Nuclear Data Center,EXFOR/CSISRS (Experimental Nuclear Reaction Data File).Database Version of November 24 (2008)